\documentclass[preprint]{aastex}

\newcommand \kms          {\hbox{km~s$^{-1}$}}
\newcommand \msol         {\hbox{M$_{\odot}$}}           

\newcommand \ha           {H$\alpha$}
\newcommand \oiii         {[\ion{O}{3}]}
\newcommand \nii          {[\ion{N}{2}]}

\begin{document}
\title{A Morphological Diagnostic for Dynamical Evolution of Wolf-Rayet Bubbles}
\author{Robert A. Gruendl, You-Hua Chu, Bryan C. Dunne and 
Sean D. Points\altaffilmark{1}}
\affil{Astronomy Department, University of Illinois, 
1002 West Green Street, Urbana, IL 61801}
\email{gruendl@astro.uiuc.edu, chu@astro.uiuc.edu,
carolan@astro.uiuc.edu,points@astro.uiuc.edu}
\altaffiltext{1}{Visiting astronomer, Cerro Tololo 
Inter-American Observatory, National Optical Astronomy Observatories, 
operated by the Association of Universities for Research in Astronomy, Inc., 
under a cooperative agreement with the National Science Foundation.}


\begin{abstract}

We have observed \ha\ and \oiii\ emission from eight of the most well
defined Wolf-Rayet ring nebulae in the Galaxy.
We find that in many cases the outermost
edge of the \oiii\ emission leads the \ha\ emission.  We suggest that
these offsets, when present, are due to the shock
from the Wolf-Rayet bubble expanding into the circumstellar envelope.
Thus, the details of the WR bubble morphology at
\ha\ and \oiii\ can then be used to better understand the physical
condition and evolutionary stage of the nebulae around Wolf-Rayet stars,
as well as place constraints on the nature of the stellar progenitor
and its mass loss history.

\end{abstract}
 

\objectname[]{G2.4+1.4}
\objectname[]{NGC 2359}
\objectname[]{NGC 6888}
\objectname[]{RCW 58}
\objectname[]{RCW 104}
\objectname[]{Sh-2 308}
\objectname[]{WR 128}
\objectname[]{WR 134}

\newpage
\section{Introduction}

Ring nebulae around Wolf-Rayet (WR) stars were first reported by 
\citet{jh65}.  The first three WR rings reported, NGC\,2359, NGC\,6888, 
and S\,308 \citep{sh59}, all show an indisputable 
shell morphology.  Since then many more ring nebulae have been identified 
in the Galaxy \citep{ch81,hbg82,mc93,metal94a,metal94b} 
and in the Magellanic Clouds \citep{cl80,detal94}.
A wide range of morphology
is observed in these WR ring nebulae, but most of these nebulae do not 
have a distinct ring morphology as did the first three identified.  While a 
small number of the WR ring nebulae may be misidentifications, in which 
the WR stars are not responsible for the nebular morphology; the
majority of them may represent the cumulative product of interactions
between the ambient interstellar medium (ISM) and the stellar winds since 
the star was formed.

To explain the wide range of morphology observed in WR rings, 
analytic models and numerical simulations have been put forward 
for wind-blown bubbles in different media: in a homogeneous ISM 
(e.g., Weaver et al. 1977), in a cloudy ISM \citep{mc84}, 
and in a circumstellar medium \citep{gsmm95a, gsmm95b}.
The most sophisticated models for WR ring nebulae have been presented
by Garc\'{\i}a-Segura et al. (1996a, b), who have taken into account
the evolution and mass loss history over the lifetime of the central star.
These models show that an interstellar bubble is formed around
a massive star 
during its main sequence lifetime and subsequently a circumstellar bubble 
will be formed interior to the interstellar bubble after the star has 
made a transition from a red supergiant (RSG) or luminous blue variable 
(LBV) to a WR star.  They further predict that evolved circumstellar 
bubbles will fragment and form ``break-out'' regions where the shock 
fronts (as traced by \oiii\ emission) lead the fragmented circumstellar 
shell material (traced by \ha\ and \nii\ emission).  

The models of Garc\'{\i}a-Segura et al. (1996a, b) are successful in 
explaining the morphology and chemical abundances of NGC\,6888 and S\,308,
but it is difficult to generalize this success to more WR rings for two 
reasons.  First, most of the WR ring nebulae have non-distinctive ring
morphology, and hence their physical nature is uncertain and comparisons 
with models can produce ambiguous results at best.  Second, most WR ring 
nebulae do not have detailed follow-up observations of morphology, 
kinematic properties, or chemical abundances for comparison with models.

To study the structure and evolution of ring nebulae around WR stars
and to compare them to Garc\'{\i}a-Segura et al.'s (1996a, b) models, 
we have selected a set of WR nebulae that have the most well-defined, 
distinct shell morphology.  These are the WR nebulae that have the 
most complete kinematic and abundance information available, and they 
are the most likely to yield meaningful results.
We have imaged these WR rings in \ha\ and \oiii$\lambda$5007 emission, 
and used the nebular morphology in these two spectral lines to study 
the physical structure of these nebulae.  We have examined their physical 
structures in conjunction with the chemical abundances, kinematic properties
and sizes of the shells, and the spectral type of the WR stars.  This 
examination is supported by comparison to models of WR ring evolution.
The results are reported in this paper.  In \S2 we describe the observations,
in \S3 we describe individual objects, and in \S4 we discuss the physical 
significance of the \oiii\ and \ha\ morphologies, and compare the observed
nebular morphologies to models and discuss their evolutionary stage.

\section{Observations and Data Reduction}

The observations were obtained with the 1~m telescope at the Mount 
Laguna Observatory (MLO) and the Curtis Schmidt Telescope at Cerro
Tololo Inter-American Observatory (CTIO) in several observing runs.
The journal of observations is given in Table~\ref{observe_tab}.

The MLO 1~m observations were made with two substantially different
detectors and narrow band filter sets.  Observations made before 1992
used a TI 800$\times$800 CCD as a detector and used a focal reducer.
The resulting images have 0\farcs 98 pixels and cover a 
13\farcm 1 field-of-view.  The typical seeing was $\sim$2\arcsec\
when these images were taken, but the focal reducer degraded the image
quality at the edge of the field to $\sim$4\arcsec.  The \ha\ filter
used had a central wavelength of 6567 \AA\ and FWHM of 62 \AA\ and thus
also transmitted the \nii $\lambda\lambda$6548, 6584 lines.  The \oiii\
observations were made with a filter centered at 5015 \AA\ with FWHM of
49 \AA, which isolated the \oiii $\lambda$5007 line.

The observations made after 1992 at MLO used a Loral 2k$\times$2k CCD.  Here
the individual frames have a 13\farcm 6 field-of-view with 
0\farcs 4 pixel$^{-1}$ scale.  Typical seeing during the observations
was between 1\farcs 5 and 2\farcs 0.  The filter used for the \ha\ 
observations has a FWHM of 20\AA\ centered at 6563 \AA\ which should
isolate the \ha\ line from most emission of the nearby \nii
$\lambda\lambda$6548, 6584 lines.  The \oiii\ filter had a FWHM of 50 \AA\
centered at 5010 \AA .  

Observations of RCW\,104 were obtained with the CTIO Curtis Schmidt on
June 29, 1992.  A Thompson 1k$\times$1k CCD detector was used
with a 31\arcmin\ field-of-view and 1\farcs 84 pixel$^{-1}$
scale.  The filter used for the \ha\ observation had a FWHM of 17 \AA\ 
centered at 6563 \AA, while the \oiii\ filter had a FWHM of 44 \AA\ 
centered at 5007 \AA .
The observations of RCW\,58 were also obtained with the Curtis
Schmidt at CTIO on December 31, 1996.  These observations used
a Tektronics 2k$\times$2k CCD detector with a 69\arcmin\ field-of-view and 
2\farcs03 pixel$^{-1}$ scale.  The filter used for the \ha\ observations
was centered at 6568 \AA\ with a FWHM of 30 \AA, while the \oiii\ 
filter had a FWHM of 40 \AA\ centered at 5023 \AA .

The observations of NGC\,6888, S\,308, WR\,128, and WR\,134 required
multiple fields to cover the entire nebula (or in the case of S\,308
to cover a portion of the object).  The individual frames have been combined
into a single mosaic for each object.  Each frame was registered to
adjacent frames using the stars in the region of overlap between frames.
Furthermore, the observing conditions often varied substantially over 
the course of the observations.  The offset due to changing sky 
background for the frames in each mosaic image were removed using the
method outlined in \citet{rg95}. 

For each final image or mosaic an astrometric solution was found for
the final image by comparison to stellar positions in the Guide Star
Catalog 1.2 \citep{ro98}.  The 
absolute positional accuracy for each image is better than 0\farcs 4.
More important for this work, the relative position accuracy of the 
\ha\ and \oiii\ images is estimated from the stellar images.  In this
case the typical uncertainty in the registration between the \ha\ and 
\oiii\ images has been reduced to less than 0.1 pixels.

\section{Results\label{sec_result}}

We detect \ha\ and \oiii\ emission from all eight WR nebulae.
In Figures 1 through 8 we present the \ha\ and \oiii\ images of these
WR ring nebulae.  In addition to the emission line images of each nebula,
we present an \oiii\ to \ha\ ratio image to show the relative position of
\oiii\ and \ha\ emission.  Note that the ratios are relative rather than 
absolute because most of the observations were made during non-photometric 
conditions.  In the \oiii/\ha\ ratio images, strong \ha\ emission (relative 
to \oiii) appears dark (black) while regions where \oiii\ emission is 
comparatively bright appear light (white).

The ratio maps (panel c in Figures 1-8) help highlight the locations
where the \ha\ and \oiii\ morphology show spatial differences.
In many cases the ratio maps clearly demonstrate that the outer edge
of the WR bubble as seen in \oiii\ emission (the \oiii\ front) is 
exterior to the edge of the bubble as traced by \ha\ emission (the
\ha\ front).  Careful examination of the \ha\ and \oiii\ images show 
four types of relationship between the \ha\ and \oiii\ fronts: \begin{itemize}
\item Type I morphology has no measurable offset between the 
\ha\ and \oiii\ fronts.  This type includes NGC\,2359 and G2.4+1.4.
\item Type II morphology has an \ha\ front trailing closely behind an 
\oiii\ front, and both fronts have a similar shape.  Examples of this 
type include WR\,128 WR\,134, S\,308, and parts of NGC\,6888 (the caps 
along the major axis).
\item Type III morphology has bright \ha\ emission trailing far behind 
a faint \oiii\ front.  In this type, the \ha\ appearance can differ
significantly from the \oiii\ appearance, for example, RCW\,58.
\item Type IV morphology has a faint \oiii\ front with no \ha\ counterpart.
For example, RCW\,104 and parts of NGC\,6888 (along the minor axis).
\end{itemize}

In Table~\ref{parm_tab} we present the physical parameters measured 
along with values taken from the literature.  There we include: spectral
type of the WR star, distance, WR wind velocity and mass loss rate, WR 
bubble radius as measured from the \oiii\ image, WR bubble expansion 
velocity, dynamical timescale ($\equiv$Radius/V$_{\rm exp}$), the
angular and linear size of typical offset between \ha\ and \oiii\ emission,
the type of offsets observed, nebular abundances, and galactocentric
distance.

\subsection{Description of Individual Nebulae}

\subsubsection{S\,308 (WR\,6)}

S\,308, around the WN5 star HD\,50896, is the closest WR ring nebula
that has been identified.  The low surface brightness and large 
angular diameter,
$\sim$40\arcmin , of this the nebula make imaging the entire structure
difficult.  Figure~\ref{s308_fig} shows the northwest and
western limbs which are also the brightest portion of the bubble.
A large, Type {\sc II}, offset between the \oiii\ and \ha\ emission
is clearly visible along the northern and western rims of S\,308.
A dramatic ``break-out'' structure is evident in the northwest corner
(best seen in the \oiii\ image, Figure~\ref{s308_fig}b).

\subsubsection{NGC\,2359 (WR\,7)}

The WR bubble in NGC\,2359 (see Figure~\ref{ngc2359_fig}) is associated 
with the the WN4 star, HD\,56925.  The \oiii/\ha\ ratio image shows no 
offset between the tracers (a Type {\sc I} morphology) along the main
shell structure, similar to the ratio maps shown by \citet{dufour94}.
In addition to
the WR bubble, there also appears a second, larger, parabolic arc of
emission which shares its eastern boundary with the WR ring nebula.
This arc of material is bounded by an HII region to the south and 
molecular material to the east and south 
\citep{schneps81,gc00}.

\subsubsection{RCW\,58 (WR\,40)}

RCW\,58 (see Figure~\ref{rcw58_fig}) is associated with the WN8 star
HD\,96548 (WR\,40).  In \ha\ emission the bubble appears to be primarily 
composed of radial filaments, while the fainter \oiii\ emission 
appears smoother and significantly more extended (Type {\sc III} morphology).
The extended \oiii\ morphology was originally noted by \citet{ch82}
and is one of the clearest offsets between the \ha\ and \oiii\
emission in our sample.  Part of the extended \oiii\ emission on the eastern
limb appears to form a concave bulge significantly different
from \ha\ morphology that may be a pronounced ``break-out'' structure.

\subsubsection{RCW\,104 (WR\,75)}

RCW\,104 (see Figure~\ref{rcw104_fig}) was originally identified by 
\citet{smith67} around the WN5 star
HD\,147419 (WR\,75).  Due to complex filamentary \ha\ emission on the same
sight line, the \ha\ emission associated with the WR star is not clearly
defined except on the western limb.   The \oiii\ emission has a 
limb--brightened ``barrel--shaped'' appearance that is
composed of tangential filaments on the eastern and western limbs.  The 
southwestern limb is the only location where both \ha\ and \oiii\ emission
are observed.  At this location, there appears to be a slight, Type {\sc II},
offset of the \oiii\ emission exterior to the \ha\ limb.  The entire 
eastern limb shows no \ha\ counterpart and therefore is classified as 
Type {\sc IV} morphology.  In the southwestern
corner the \oiii\ emission has a faint structure similar in character to 
the ``break--out'' structure in RCW\,58 and S\,308.

\subsubsection{G2.4$+$1.4 (WR\,102)}

G2.4$+$1.4 (see Figure~\ref{wr102_fig}) is another ring nebulae in our
sample that has no discernible offset between \oiii\ and \ha\ (Type 
{\sc I} morphology).  Among the WR stars in our sample, WR\,102, or 
LSS\,4368, is the only WO star.
The field around WR\,102 is highly confused by interstellar emission along
the same sight line.  \citet{dl90} have made extensive kinematic 
observations and suggested that the bubble is a ``blister'' structure
on the front of a dense cloud.

\subsubsection{Anon (WR\,128)}

Previous \ha\ and \oiii\ images of nebulosity associated with WR\,128 
have concentrated on the two prominent arcs to the north and east of 
the central WN4 star, HD\,187282.  There are no published emission 
line images that extend an equal distance to the south and west to 
determine whether the partial ring has other components.  
We have imaged an area that extends as far 
south and west as to the north and east and and confirm that there 
is no bright \ha\ or \oiii\ emission on the opposite side of HD\,187282.
Our images (Figure~\ref{wr128_fig}) do show that the arc east of 
HD\,187282 continues further south than previously known \citep{mc93}.
The composite image (Figure~\ref{wr128_fig}c) shows that the \oiii\
emission is significantly offset exterior from the \ha\ emission with
respect to the central star (Type {\sc II} morphology).
Recent HI observations by \citet{arnal99}
show that there is an \ion{H}{1} hole coincident with WR\,128.  
In the \ha\ image the arc of ``limb-brightened''
emission appears to be ``filled'' with low surface brightness emission
from the shell to roughly the position of the WR star.

\subsubsection{Anon (WR\,134)}

WR\,134 (see Figure~\ref{wr134_fig}) is the WN6 star HD\,191765. The nebular
images show markedly different morphology in \ha\ and \oiii .  The \ha\ 
image shows a hemisphere of filamentary emission along with non-uniform
diffuse \ha\ emission.  The \oiii\ emission is all filamentary and is
only present to the northwest of HD\,191765.  Much of the \ha\ emission
southwest of the WR star has been shown by \citet{tc82} to be kinematically 
distinct from the filamentary emission. The \oiii\ filaments 
appear slightly offset radially outward with respect to the \ha\ filaments
(Type {\sc II} morphology).
The WR star is projected close to the center of the filamentary
emission.  In the composite image, offsets between
the filaments in \ha\ and \oiii\ are not very clear, in part due to the
the large dynamic range in the \ha\ image.
The outermost \oiii\ filaments are similar in morphology to the ``break-out''
features observed in NGC\,6888.

\subsubsection{NGC\,6888 (WR\,136)}

NGC\,6888 around the WN6 star HD\,192163 is often regarded as the 
proto-typical WR bubble.  The \ha\ and \oiii\ images 
(see Figure~\ref{ngc6888_fig}) shows a clear offset between the
\ha\ and \oiii\ emission along the limb-brightened shell.
Furthermore, a faint network of filamentary structures appear to envelope
the \ha\ emission (Type {\sc II} morphology). 
Beside the filamentary structures near the \ha\ emission,
there is a large \oiii\ front with no \ha\ counterpart that bulges
outward from the northwest limb of the bubble near the minor axis
(Type {\sc IV} morphology).
This was previously noted
by \citet{mc93} and by \citet{dufour94}.
In Figure~\ref{ngc6888_fig}c it is apparent that the total extent
of the \oiii\ emission is centered on the WR star, while the limb-brightened
\ha\ shell is off-center.  High resolution Hubble Space Telescope 
{\it WFPC2} images of a portion of the shell rim presented
by \citet{mhc00} show the offsets along the shell rim with dramatic
clarity.

\section{Discussion}

\subsection{Physical Significance of \ha\ and \oiii\ Morphologies
\label{sec_phys} }

\ha\ and \oiii\ line images of a nebula provide useful diagnostics 
of its physical structure.  \ha\ is a recombination line and thus
shows the overall distribution of ionized material, but because
the line strength of \ha\ decreases with temperature its sensitivity
drops for temperatures $>$10$^4$ K.
On the other hand, the \oiii\ line originates in a forbidden 
transition whose upper level is populated by collisional 
excitation; therefore, its intensity increases with temperature, 
and \oiii\ line images trace high excitation regions with 
temperatures $\ge10^4$ K.
Consequently, we expect the \ha\ and \oiii\ morphologies to be 
different, especially over regions where physical conditions 
change rapidly.

Behind a shock front, radiative cooling takes place; as the 
temperature drops, the density increases.  Thus, a displacement 
between \oiii\ emission and \ha\ emission occurs \citep{cox72}.
The magnitude of the displacement depends on the post-shock 
temperature, which is determined by the shock velocity, and the 
cooling rate, which is dependent on the density and metallicity. 
In cases where a shock propagates into a dense medium, the
cooling rate behind the shock front is high, the cooling zone
is narrow, and the offset between \ha\ and \oiii\ emission peaks
is minimal.  
For a shock propagating in a tenuous medium, the cooling rate is
lower, the cooling zone is wider, thus the offset between the
\ha\ and \oiii\ emission peaks is larger and may be observable.
If the ambient density is low enough, it is possible that 
\oiii\ emission is observable behind the shock, while \ha\ 
emission is too faint to be detected.
Finally, if a shock propagates through a high-density medium and
then into a low-density medium, it is possible to see bright \ha\
emission associated with the dense medium and a leading \oiii\
front associated the recently shocked low-density medium.
This could produce the largest displacement between the leading 
edges of \ha\ and \oiii\ emission.

\citet{cox72} has calculated the expected 1-d surface brightness
profile for H$\beta$ and \oiii\ assuming a planar shock propagating
at 100 \kms\ into a medium with density of 1 cm$^{-3}$ (see Figure 3
in Cox 1972).  This calculation found the separation between the 
peak H$\beta$ and \oiii\ emission to be roughly 0.001 pc. 
For a WR star that had a RSG progenitor, its WR bubble will be
expanding into a circumstellar envelope whose density can be 
calculated from the RSG wind properties.  For a RSG mass loss
rate of 10$^{-5}$ M$_\odot$~yr$^{-1}$ and a wind velocity of
50 km~s$^{-1}$, the RSG wind density at 6 pc from the star
would be $\sim$0.02 cm$^{-3}$.  If we assume that the shock from the WR 
bubble expanding into the circumstellar envelope (RSG wind) at
100 \kms, then the post-shock density would be $\sim$50 
times lower and the cooling timescale $\sim$50 times longer
that Cox's results.  
Thus, the separation between H$\alpha$ (or H$\beta$) and \oiii\
would be $\sim$50 times greater, or $\sim$0.05 pc.  This order of 
magnitude estimate is similar to the typical observed separation
between \ha\ and \oiii\ fronts in our sample (see Table~\ref{parm_tab}).

The four types of relationship between the \ha\ and \oiii\ fronts
described in \S 3 can all be interpreted as shocks propagating into 
different ambient media.   WR bubbles with the Type I \ha\ and \oiii\
fronts may be expanding into dense media, the cooling rate is high, 
and the displacement between \ha\ and \oiii\ emission peaks cannot be
resolved by our images.
WR bubbles with Type II \ha\ and \oiii\ fronts are probably 
expanding into a medium with density low enough that the cooling
time behind the shock is long, and therefore an observable 
displacement between the \oiii\ and \ha\ fronts is produced.
Type III \ha\ and \oiii\ fronts can be produced if a shock has 
propagated through a dense medium into a tenuous medium.  The \ha\
morphology is determined by the dense medium, while the \oiii\ 
emission traces the progression of the shock in the tenuous 
medium.  WR bubbles with Type {\sc IV} morphology are expanding into a 
very tenuous medium.  The post-shock density is low and
temperature is high, so that only \oiii\ emission, but not \ha,
is strong enough to be detected.

\subsection{Comparison with Predictions of Theoretical Models}

Garc\'{\i}a-Segura et al. (1996a,b) have simulated the hydrodynamical 
evolution of circumstellar gas around a massive star, taking into account
the stellar evolution and mass loss history starting from the main sequence
through the WR phase.  WR stars can be descendants of either LBV or 
RSG progenitors; accordingly, Garc\'{\i}a-Segura et al. have simulated
nebulae around a 60 \msol\ star, which has gone through an LBV phase,
and a 35 \msol\ star, which has gone through a RSG phase.
These simulations produce WR bubbles with distinctly different morphologies,
kinematics, and abundances.  These predictions, when compared with 
observations, can be used to determine the evolutionary stage of the
WR bubble and the nature of the WR progenitor.
 
For a 60 \msol\ star, \citet{gsmml96} predict that the resulting WR bubble
will be observable only in the early WR stage.  
After the WR bubble has swept up most of the LBV wind material, it will expand
into a low density circumstellar medium.  This drop in the ambient density,
causes an increase in the shock velocity and the fragmentation of the 
dense WR shell.  Thus, the resultant observable WR bubble will have a
radial filamentary \ha\ structure in the dense shell and will be surrounded 
by \oiii\ emission behind the accelerating outer shock in the tenuous 
circumstellar medium.  This fits precisely the Type III \ha\ and \oiii\
morphology observed in WR bubbles.  Indeed \citet{gsmml96} have suggested
that RCW\,58 was produced this way and that its central star is a 
descendant of a star that went through an LBV phase.

For a 35 \msol\ star, \citet{gslmm96} have performed calculations for stars
with RSG wind velocities of 15 and 75 \kms.  For the case of a 75 \kms\ 
RSG wind velocity, they predict that the swept-up WR shell is dense enough 
to be observable only during the early stages, when the radius is $\le$ 1 pc.  
For the case of a RSG with a slow wind, $\sim$15 \kms, their model shows
that the outer edge of the RSG wind is compressed into a RSG shell and
the inner edge is swept up by WR wind into a WR shell.  The WR shell
collides with the outer RSG shell after $\sim$10$^4$ years, leading to 
a deceleration and brightening of the WR shell.
After the collision, the dense clumps in the WR shell, having higher 
inertia, cross the RSG shell and form blow-outs.  Eventually, the 
whole bubble fragments and the shocked WR wind breaks out and forms 
an outer shock leading the fragmented WR shell.

Note that these models can be used only qualitatively as many 
approximations/simplifications have been made.  We note particularly 
that the physical distinction between the models using 15 and 75 \kms\ 
RSG wind is really in the density of the RSG wind, which is proportional
to $\dot{\rm M}/$V, where $\dot{\rm M}$ is the mass loss rate of the
RSG and V is the RSG wind velocity.  Therefore, the model of \citet{gslmm96}
for the slow RSG wind is applicable for a faster RSG wind with a larger
RSG mass loss rate.

Garc\'{\i}a-Segura et al.'s (1996a,b) model for a low-density RSG wind
(low $\dot{\rm M}/$V) produce WR bubbles that have thick shells when
they are dense enough to be observable in \ha\ and \oiii.  
None of the WR bubbles we observe fit this description.  On the other
hand, their model for a high-density RSG wind (high $\dot{\rm M}/$V)
produce WR bubbles with thin shells that are most observable when the
WR shell collides with the RSG shell.  
Within 1,000--2,000 years after the collision, the thin shell expands into
a tenuous circumstellar medium and its \oiii\ emission should lead 
the \ha\ emission; similar to the Type II morphology.  After the 
WR shell fragments and breakouts occur, the outer shocks may be 
observable in \oiii\ but not in \ha, producing a Type IV morphology.

Garc\'{\i}a-Segura et al. (1996a,b) suggested that NGC\,6888 is produced
by a WR star which went through its RSG phase with a slow wind velocity, 
and S\,308 with a faster RSG wind velocity.  Our images support
their interpretation that NGC\,6888 had a denser RSG wind than S\,308.
S\,308 must be near the stage where the WR shell has collided with the
RSG shell.  The thickness and break-out structure of the S\,308 shell
are more consistent with the predictions of their slow RSG wind model
(high $\dot{\rm M}/$V).  We suggest that NGC\,6888, S\,308, WR\,128, 
WR\,134, and RCW\,104 all have RSG progenitors, and that the RSG wind 
of NGC\,6888 was significantly denser than the others.\footnote{Abundance
measurements show enrichment of CNO processed elements in NGC\,6888, 
S\,308, and RCW\,104, which demonstrates the circumstellar origin of
the bubble material.  The abundance measurements of the nebula around
WR\,134 were made in the diffuse \ha\ arc \citep{es92}, which is 
kinematically different from the \oiii\ bubble and has been
identified as an HII region \citep{tc82}.  Future abundance measurements
need to be made for the nebula around WR\,128 and the \oiii\ filaments
around WR\,134 to confirm their circumstellar origin.}

Finally, two WR bubbles in our sample do not resemble any of 
the bubble morphologies expected in Garc\'{\i}a-Segura et al.'s (1996a,b)
models.  These two bubbles, NGC\,2359 and G2.4$+$1.4, show no offsets
between the \ha\ and \oiii\ fronts.  These two objects are also
the smallest and have the slowest expansion velocities.  When compared 
with the other WR bubbles in our sample, their dynamical timescales are 
long (old) for their size.  These results may be explained by their
interstellar environment, which is denser than what Garc\'{\i}a-Segura et 
al. (1996a,b) have assumed.  NGC\,2359, being adjacent to molecular clouds,
is clearly in a denser interstellar environment.  Consequently, the
interstellar bubble blown by the progenitor over its main sequence 
lifetime is small and dense, and the circumstellar bubble of the WR
star can easily merge with this interstellar bubble.  The bubble
in NGC\,2359 is dominated by swept-up interstellar material, as indicated 
by the abundances measured \citep{es92}.  The situation for G2.4$+$1.4 is 
less certain because of its confusing line of sight near the Galactic 
center \citep{dl90}.

\subsection{Future Work}

The analysis presented in this paper has been unavoidably qualitative.
A number of observations are needed to better understand the physical
state of circumstellar material around WR star so that a quantitative
assessment can be made.  Due to the low density in the bubbles, 
flux-calibrated narrow-band imaging in the \ha\ line would yield 
detailed surface brightness profiles of the gas and allow us to derive
the rms density of the bubble.  
High-dispersion spectroscopic observations of \ha\ and \oiii\ would 
show the relative expansion velocity of shock fronts in each tracer
and should verify our interpretation of Type {\sc III} and {\sc IV}
morphologies.  
Furthermore, these observations would provide a detailed
snapshot of the kinematics of filaments and knots in the fragmenting
shell that might identify the instabilities responsible for their formation.
Detailed abundance measurements within the shock
fronts, filaments, and knots will probe not only the chemical composition
but also the nucleosynthetic yields and mixing as a massive star evolves.
Finally, sensitive X-ray observations with new instruments such as 
{\em XMM-Newton} would probe the shocked fast wind that drives the nebular
expansion.  Potentially, these observations could probe the mass loss 
history of the central star, pinpoint the nature of the progenitor, and 
make a detailed assessment of the physical processes that occur in 
bubble formation.

\acknowledgements
R.A.G. is supported by the NASA grant SAO GO 0-1004X.
The use and operation of the Mount Laguna Observatory are supported by the 
Astronomy Department of the University of Illinois and the Astronomy 
Department of San Diego State University.

\clearpage

\begin{figure}
\figcaption{Images of S\,308. Panel (a) shows \ha\ emission, panel (b)
shows \oiii\ emission for the same field of view as (a), and panel (c)
shows a composite of \ha\ (black) and \oiii\ (white) emission.  For more
details about the composite image see Section~\ref{sec_result}). A 
cross marks the WR star associated with the nebulosity in each image. 
\label{s308_fig}}
\end{figure}

\begin{figure}
\figcaption{Images of NGC\,2359, see Figure~\ref{s308_fig} for details.
\label{ngc2359_fig}}
\end{figure}

\begin{figure}
\figcaption{Images of RCW\,58, see Figure~\ref{s308_fig} for details.
\label{rcw58_fig}}
\end{figure}

\begin{figure}
\figcaption[rcw104_3pan.eps]{Images of RCW\,104, see 
Figure~\ref{s308_fig} for details.  \label{rcw104_fig}}
\end{figure}

\begin{figure}
\figcaption{Images of G2.4$+$1.4, see Figure~\ref{s308_fig} for details.
\label{wr102_fig}}
\end{figure}

\begin{figure}
\figcaption{Images of WR\,128, see Figure~\ref{s308_fig} for details.
\label{wr128_fig}}
\end{figure}

\begin{figure}
\figcaption{Images of WR\,134, see Figure~\ref{s308_fig} for details.
\label{wr134_fig}}
\end{figure}

\begin{figure}
\figcaption{Images of NGC\,6888, see Figure~\ref{s308_fig} for details.
\label{ngc6888_fig}}
\end{figure}

\clearpage
\begin{deluxetable}{llllcc}
\tablewidth{0pt}
\tablecaption{Observation Summary\label{observe_tab}}
\tablehead{
\colhead{Object}  & \colhead{Telescope} & \colhead{Date} & 
\colhead{Filter} & \colhead{Exposure}   & \colhead{Resolution} \\
\colhead{ }       & \colhead{}          & \colhead{}     &
\colhead{[\AA ]\tablenotemark{a}} & \colhead{[sec]}     & \colhead{[\arcsec ]} 
}
\startdata
NGC\,2359  & MLO 1m          & Oct.  1999 & 6563/20 &  1800 & 3.5 \\
           & MLO 1m          & Oct.  1999 & 5015/49 &  1800 & 4.0 \\
G2.4$+$1.4 & MLO 1m          & July  1991 & 6567/62 &   600 & 3.3 \\
           & MLO 1m          & July  1991 & 5015/49 &   600 & 3.3 \\
WR\,128    & MLO 1m          & July  1999 & 6563/20 &  1200 & 3.0 \\
           & MLO 1m          & July  1999 & 5015/49 &  1200 & 2.3 \\
WR\,134    & MLO 1m          & July  1991 & 6567/62 &   600 & 2.9 \\
           & MLO 1m          & July  1991 & 5015/49 &   600 & 3.0 \\
NGC\,6888  & MLO 1m          & July  1991 & 6567/62 &   300 & 3.1 \\
           & MLO 1m          & July  1991 & 5015/49 &   900 & 3.6 \\
S\,308     & MLO 1m          & Jan.  1999 & 6563/20 &  1200 & 3.5 \\
           & MLO 1m          & Jan.  1999 & 6563/20 &  1200 & 3.4 \\
RCW\,58    & Curtis--Schmidt & Dec.  1996 & 6568/30 &   900 & 3.8 \\
           & Curtis--Schmidt & Dec.  1996 & 5023/40 &  1800 & 3.4 \\
RCW\,104   & Curtis--Schmidt & June  1992 & 6563/14 &   600 & 2.6 \\
           & Curtis--Schmidt & June  1992 & 5007/44 &   600 & 2.9 \\
\enddata
\tablenotetext{a}{Values are given as central wavelength/FWHM}
\end{deluxetable}

\clearpage

\begin{deluxetable}{rrrrrrrrr}
\tablewidth{0pt}
\rotate
\scriptsize
\tablecaption{Observed and Measured Parameters\label{parm_tab}}
\tablehead{
\colhead{}          & \multicolumn{8}{c}{Object} \\
\colhead{Quantity} & {NGC\,2359} & {G2.4$+$1.4} &  {WR\,128} & {WR\,134} & {NGC\,6888} &  {RCW\,58}   & {RCW\,104} & {S\,308}     
}
\startdata
Spectral Type                       &      WN4 &     WO1 &      WN4 &      WN6 &     WN6 &      WN8 &     WN5 &     WN5 \\
Distance~[kpc]\tablenotemark{a}     &      3.5 &     4.0 &      4.0 &      2.1 &     1.8 &      4.0 &     4.0 &     1.6 \\
V$_{\infty}$~[\kms]\tablenotemark{b}&  \nodata & \nodata &     2270 &     1905 &    1605 &      975 & \nodata &    1720 \\
$\log{(\dot{\rm M }~[{\rm M}_\odot~{\rm yr}^{-1}])}$\tablenotemark{b} &  
                                         \nodata & \nodata &  \nodata &  $-$4.13 & $-$4.02 &  \nodata & \nodata & $-$4.12 \\
Radius~[pc]\tablenotemark{c}        &     2.1  &     2.7 &      6.2 &      6.6 &     4.5 &      5.7 &     5.9 &     9.0 \\
V$_{\rm exp}$~[\kms ]\tablenotemark{d} &    18 &      42 &  \nodata &    $>$50 &      80 &      110 & \nodata &      65 \\
Dyn. Timescale [10$^4$ yr]          &     11.3 &     8.8 &  \nodata &  $<$10.7 &     5.5 &      5.1 & \nodata &    13.5 \\
Typical Offset [$^{\prime\prime}$]  &   $<$1.5 &    $<$1 &     3--5 &     4--8 &      10 &    18-40 &     3.0 &  16--20 \\
Typical Offset [pc]                 & $<$0.025 & $<$0.03 &0.06--0.10&0.04--0.08&    0.09 &0.35--0.78&    0.06 & 0.12--0.17\\
Offset Type                         &  {\sc I} & {\sc I} & {\sc II} &{\sc II} &{\sc II,IV}&{\sc III}&{\sc II,IV}& {\sc II}\\
$[12+\log{O/H}]$\tablenotemark{e}   &     8.24 &    8.45 &  \nodata &     8.83 &    8.14 &     8.72 &    8.48 &    8.03 \\
$[\log{N/O}]$\tablenotemark{e}      &  $-$0.95 & $-$0.49 &  \nodata &  $-$0.85 & $+$0.27 &  $-$0.30 & $-$0.13 & $+$0.22 \\
Y\tablenotemark{e}                  &      0.3 &    0.47 &  \nodata &     0.31 &    0.29 &     0.43 &    0.31 &    0.45 \\
Gal. Distance~[kpc]\tablenotemark{a} &     15 &       7 &        7 &       10 &      10 &       10 &       8 &      11 \\
\enddata

\tablenotetext{a}{Distance and galactocetric distance are from \citet{vdh88}.}
\tablenotetext{b}{Stellar Wind V$_{\infty}$ and $\log{(\dot{\rm M })}$ from \citet{pr90}.}
\tablenotetext{c}{Measured from the \oiii\ image.}
\tablenotetext{d}{Expansion velocities are from the following: NGC\,2359, WR\,134, and NGC\,6888 -- \citet{tc82};
G2.4$+$1.4 -- \citet{dl90}; RCW\,58 -- \citet{ch88}; S\,308 -- \citet{ch82b}.}
\tablenotetext{e}{Abundance Measurements from \citet{es92}; Local ISM Abundances: 12$+\log{O/H}=$8.7, $\log{N/O}=-$1.12, Y$=$0.285}
\end{deluxetable}


\clearpage

\begin{thebibliography}{}

\bibitem[Arnal et al. (1999)]{arnal99} Arnal, E. M., Cappa, C. E., Rizzo, J. R.,
\& Cichowolski, S.\ 1999, \aj, 118, 1798 


\bibitem[Chu (1981)]{ch81} Chu, Y.-H.\ 1981, \apj, 249, 195

\bibitem[Chu (1982)]{ch82} Chu, Y.-H.\ 1982, \apj, 254, 578

\bibitem[Chu (1988)]{ch88} Chu, Y.-H.\ 1988, \pasp, 100, 986 
 
\bibitem[Chu et al. (1982)]{ch82b} Chu, Y.-H., Gull, T. R., Treffers, R. R., 
Kwitter, K. B., and Troland, T. H., 1982, \apj, 254, 562 

\bibitem[Chu \& Lasker (1980)]{cl80} Chu, Y.-H., \& Lasker, B. M.\
1980, \pasp, 92, 703


\bibitem[Cox (1972)]{cox72} Cox, D. P. 1972, \apj, 178, 143

\bibitem[Dopita et al. (1994)]{detal94} Dopita, M. A., Bell, J. F., 
Chu, Y.-H., \& Lozinskaya, T. A.\ 1994, \apjs, 93, 455

\bibitem[Dopita \& Lozinskaya (1990)]{dl90} Dopita, M. A., \& 
Lozinskaya, T. A., 1990, \apj, 359, 419

\bibitem[Dufour (1994)]{dufour94} Dufour, R. J.\ 1994, in Circumstellar
Media in Late Stages of Stellar Evolution, eds. R. E. S. Clegg, 
I. R. Stevens, \& W. P. S. Meikle, (Cambridge:University Press), 78

\bibitem[Esteban et al. (1992)]{es92} Esteban, C., Vilchez, J. M., 
Smith, L. J., \& Clegg, R. E. S.\ 1992, \aap, 259, 629

\bibitem[Garc\'{\i}a-Segura \& Mac Low (1995a)]{gsmm95a} Garc\'{\i}a-Segura, G.
\& Mac Low, M.-M.\ 1995a, \apj , 455, 145

\bibitem[Garc\'{\i}a-Segura \& Mac Low (1995b)]{gsmm95b} Garc\'{\i}a-Segura, G.
\& Mac Low, M.-M.\ 1995b, \apj , 455, 160

\bibitem[Garc\'{\i}a-Segura et al. (1996a)]{gsmml96} 
Garc\'{\i}a-Segura, G., Mac Low, M.-M., \& Langer, N.\ 1996a, \aap, 305, 229

\bibitem[Garc\'{\i}a-Segura et al. (1996b)]{gslmm96}
Garc\'{\i}a-Segura, G., Langer, N., \& Mac Low, M.-M.\ 1996b, \aap, 316, 133  

\bibitem[Gruendl \& Chu (2000)]{gc00} Gruendl, R. A., \& Chu, Y.-H.
2000, in preparation (to be submitted to the AJ)

\bibitem[Heckathorn et al. (1982)]{hbg82} Heckathorn, J. N., Bruhweiler, 
F. C., \& Gull, T. R.\ 1982, \apj, 252, 230

\bibitem[Johnson \& Hogg (1965)]{jh65} Johnson, H. M., \& Hogg, D. E.\ 1965, 
\apj , 142, 1033

\bibitem[Marston et al. (1994a)]{metal94a} Marston, A. P., Chu, Y.-H., \& 
Garc\'{\i}a-Segura, G.\ 1994a, \apjs, 93, 229

\bibitem[Marston et al. (1994b)]{metal94b} Marston, A. P., Yocum, D. R.,
Garc\'{\i}a-Segura, G., \& Chu, Y.-H.\ 1994b, \apjs, 95, 151

\bibitem[McKee, van Buren, \& Lazareff (1984)]{mc84} McKee, C.F., van Buren, 
D., \& Lazareff, B.\ 1984, \apjl, 278, L115

\bibitem[Miller \& Chu (1993)]{mc93} Miller, G. J., \& Chu, Y.-H. 1993,
\apjs, 85, 137

\bibitem[Moore, Hester, \& Scowen (2000)]{mhc00} Moore, B. D., Hester, J. J.,
\& Scowen, P. A. 2000, \aj, in press (astro-ph/0003053)


\bibitem[Prinja et al. (1990)]{pr90} Prinja, R. K., Barlow, M. J., \&
Howarth, I. D.\ 1990, \apj, 361, 607

\bibitem[Regan \& Gruendl (1995)]{rg95} Regan, M. W. \& Gruendl, R. A.\
1995, in ADASS IV Proceedings, ASP Conference Series 77, eds. R. A. Shaw, H. E.,
Payne, \& J. J. E. Hayes, 335 

\bibitem[Roser et al. (1998)]{ro98} Roser, S., Morrison, J. E., Bucciarelli, B.,
Lasker, B., \& McLean, B. J.\ 1998, in IAU Symposium 179, New Horizons from
Multi-wavelength Sky Surveys, eds B. J. McLean, D. A. Golombek, J. J. E. Hayes,
and H. E. Payne (Dordrecht:Kluwer), 420

\bibitem[Schneps et al. (1981)]{schneps81} Schneps, M. H., Hascheck, A. D., 
Wright, E. L., \& Barrett, A. H.\ 1981, \apj , 243, 184.

\bibitem[Sharpless (1959)]{sh59} Sharpless, S. 1959, \apjs, 4, 257

\bibitem[Smith (1967)]{smith67} Smith, L. J.\ 1967, Ph.D. Thesis, Australian
National University 

\bibitem[Treffers \& Chu (1982)]{tc82} Treffers, R. R., \& Chu, Y.-H., 1982,
\apj, 254, 569 

\bibitem[van der Hucht et al. (1988)]{vdh88} van der Hucht, K. A., 
Hidayat, B., Admiranto, A. G., Supelli, K. R., \& Doom, C.\ 1988, 
\aap, 199, 217

\bibitem[Weaver et al. (1977)]{w77} Weaver, R., McCray, R., Castor, J.,
Shapiro, P., \& Moore, R.\ 1977, \apj , 218, 377


\end{thebibliography}
\end{document}